# Beyond freezing: amorphous water in biomimetic soft nanoconfinement

Livia Salvati Manni[1,2†], Salvatore Assenza[2†], Michael Duss[1], Jijo J. Vallooran[1,2], Fanni Juranyi[3], Simon Jurt[1], Oliver Zerbe[1], Ehud M. Landau[1*], Raffaele Mezzenga[2,4*]

[1] Department of Chemistry, University of Zürich, Winterthurerstrasse 190, 8057 Zürich (Switzerland)

[2] Department of Health Sciences & Technology, ETH Zürich, Schmelzbergstrasse 9, 8092 Zürich (Switzerland)

[3] Laboratory for Neutron Scattering, Paul Scherrer Institut, CH-5232 Villigen PSI

[4] Department of Materials, ETH Zurich, Wolfgang-Pauli-Strasse 10, 8093 Zurich, (Switzerland)

*Water is a ubiquitous liquid with unique physico-chemical properties, whose nature has shaped our planet and life as we know it. Water in restricted geometries has different properties than in bulk. Confinement can prevent low-temperature crystallization into a hexagonal structure, thus creating a state of amorphous water. In this work we introduce a family of synthetic lipids with designed cyclopropyl modification in the hydrophobic chains that exhibit unique liquid-crystalline behaviour at low temperature, enabling maintenance of amorphous water down to ≈10 K due to nanoconfinement in a bio-mimetic milieu. Small and Wide Angle X-ray Scattering, Elastic and Inelastic Neutron Scattering, Nuclear Magnetic Resonance Spectroscopy and Differential Scanning Calorimetry, complemented by Molecular Dynamics Simulations, unveil a complex lipid/water phase diagram, in which bicontinuous cubic and lamellar liquid crystalline phases containing sub-zero liquid, glassy, or ice water emerge as a competition between the two components, each pushing towards its thermodynamically favoured state.*

Physical confinement of water at the nanoscale can play a major role in controlling its properties, with fundamental implications in physical, chemical, geological and biological phenomena.[1,2] Not surprisingly, the mobility of nanoconfined water along with its behaviour at interfaces has attracted widespread attention in past and recent times.[3-5] In this regard, the nature of the interface and the geometric details of the confining surface are key parameters.[6,7] In particular, confinement in the nanometre range can inhibit the arrangement of water molecules into an ice structure, thereby preventing crystallization at sub-zero temperature and creating a state of amorphous water.[8-10]

Confinement within soft interfaces, such as those formed by the self-assembly of surfactants in aqueous environment, has been suggested as a model for confined water in a cellular environment.[6,11-13] Thus understanding the properties of water in such systems can provide key insights to the mechanisms of cell survival at low temperatures.[14] In particular, confinement effects have been reported in various phases formed by hydrated monoacylglycerols.[15,16] The polymorphism of the most commonly studied monoacylglycerols at different hydration levels and temperatures includes lamellar ($L_\alpha$), inverse bicontinuous cubic ($V_2$), inverse hexagonal ($H_{II}$) and inverse micellar ($L_2$) liquid crystalline phases.[17,18] Specifically, the $L_\alpha$ phase consists of stacked planar lipidic bilayers separated by slabs of water, where the lipid tails have no positional order within a single layer, while the $V_2$ phases comprise a single curved lipidic bilayer arranged in three-dimensional space along continuous periodic minimal surfaces. Three distinct $V_2$ phases can be formed according to the surface symmetry: gyroid (*Ia3d*), diamond (*Pn3m*), or primitive (*Im3m*), and in all cases, the lipidic bilayer is surrounded by two identical, but not interconnected water channels.[19] In $H_{II}$, the lipids aggregate into inverse micellar cylinders that are packed into a hexagonal lattice.[20] Among these geometries, cubic phases are of particular interest due to their unique properties, including solid gel-like consistency, transparency, high interfacial surface area, and thermodynamic stability in excess water.[21] These features, together with their ability to solubilize hydrophilic and hydrophobic guest molecules, render these biomaterials useful in various applications ranging from biosensors and drug delivery systems[22] to catalysis.[23] Most remarkably, cubic phases are ideal matrices for reconstitution,[24] stabilization and crystallization of membrane proteins.[25,26]

Unfortunately, just below room temperature, the rich monoacylglycerols polymorphism is lost, as this common class of lipids crystallizes into a lamellar crystalline phase ($L_c$), in which the lipidic tails pack into a crystalline lattice that exhibits long-range order.[27] Furthermore, below 0 °C a coexistence of $L_c$ with ice is found at all hydrations.[17] This not only limits any practical use of lipidic mesophases at low temperature, but also prevents using these systems to study the properties of sub-zero nanoconfined water, due to its occurrence as crystalline ice. Recently,[28] however, we have shown that by simply replacing the *cis* double bond in the middle of the lipidic chain of the most studied monoacylglycerol, Monoolein (MO), with a cyclopropyl moiety, a novel lipid monodihydrosterculin (MDS) is obtained, whose phase behaviour reveals the absence of the reverse hexagonal phase and the stability of the Pn3m phase down to 4°C. This feature is enough to facilitate, for example, membrane protein reconstitution and crystallization at temperatures as low as 4°C, where standard monoacylglycerols are frozen into the lamellar crystalline phase $L_c$.[28] The atypical behaviour of hydrated MDS at low temperature is reminiscent of the ability of cyclopropanated phospholipids to enhance bacterial resistance to cold shock,[29] suggesting an effect of the cyclopropanation in stabilizing lipidic liquid crystalline phases at sub-zero temperatures.

Inspired by this favourable background and in order to investigate the phase properties of novel lipids with restricted geometries, we here introduce the design of a library of cyclopropyl-modified monoacylglycerols, in which the rigidity of the lipidic tail can be modulated by changing the number and the position of the cyclopropyl group, along with the length and the curvature of the hydrophobic chains. In this way, a new family of designer lipids is presented that form mesophases with unconventional properties at low temperature, the most remarkable being the capacity to confine glassy water down to at least 10 K, even with extremely slow cooling rates. To the best of our knowledge, our study shows for the first time the presence of amorphous water in soft biomimetic interfaces at such low temperature and in quasi-equilibrium conditions.

Through a systematic study, based on Small Angle X-ray Scattering (SAXS), the role of the rigidity of the lipid tail on the ensuing phase behaviour is elucidated. We exemplify the versatility and the power of the approach on two cyclopropanated monoacylglycerols, taking their *cis* double-bonded homologues as a benchmark. Particular attention is given to the low temperature phase behaviour of dicyclopropylmonolinolein (DCPML). In this system, the thermal behaviour of the confined water is investigated using Differential Scanning Calorimetry (DSC), Elastic and Inelastic Fixed Window Scan Neutron Scattering (FWS), Wide Angle X-ray Scattering (WAXS) and Nuclear Magnetic Resonance (NMR), revealing a system with unique low temperature behaviour due to a subtle lipid/water interplay. This was complemented by Molecular Dynamics simulations, which allowed interpreting and addressing ice stability and water mobility within the lipid confined geometries.

**Lipidic polymorphism**

To date, the number of natural lipids that form $V_2$ phases is limited, and most systems consist of hydrated isoprenoid lipids or unsaturated monoacylglycerols. New synthetic lipids have been designed and synthesized by varying the head group,[30] but no studies report the effect of systematic synthetic modifications in the hydrophobic chain region on the phase behaviour. The lipid chain provides the scaffolding element of all mesophases: its molecular structure, the specific length and position and degree of unsaturation, as well as curvature, greatly affect the structural properties of the ensuing mesophases.[18,31]

Replacement of the *cis* double bond of monoacylglycerols with a chemically analogous *cis* cyclopropyl moiety maintains a similar chain curvature and lipid length, but results in a substantial variation of packing frustration and lateral stress of the tails. The rigidity of the lipidic tail can be further modulated by changing the number and the position of the cyclopropyl groups, as well as the length and the curvature of the hydrophobic chains. In addition to MDS,[28] the cyclopropanated lipids monolactobacillin (MLB) and dicyclopropylmonolinolein (DCPML), analogues of monolinolein (ML) and monovaccein (MV), respectively, were synthesized (see SI) (**Fig. 1a**). The full binary phase diagrams for these designer cyclopropanated monoacylglycerols were established by SAXS, and are shown in **Fig. 1b, 1c,** overlaid on the corresponding phase diagrams of the olefinic monoacylglycerols.[18]

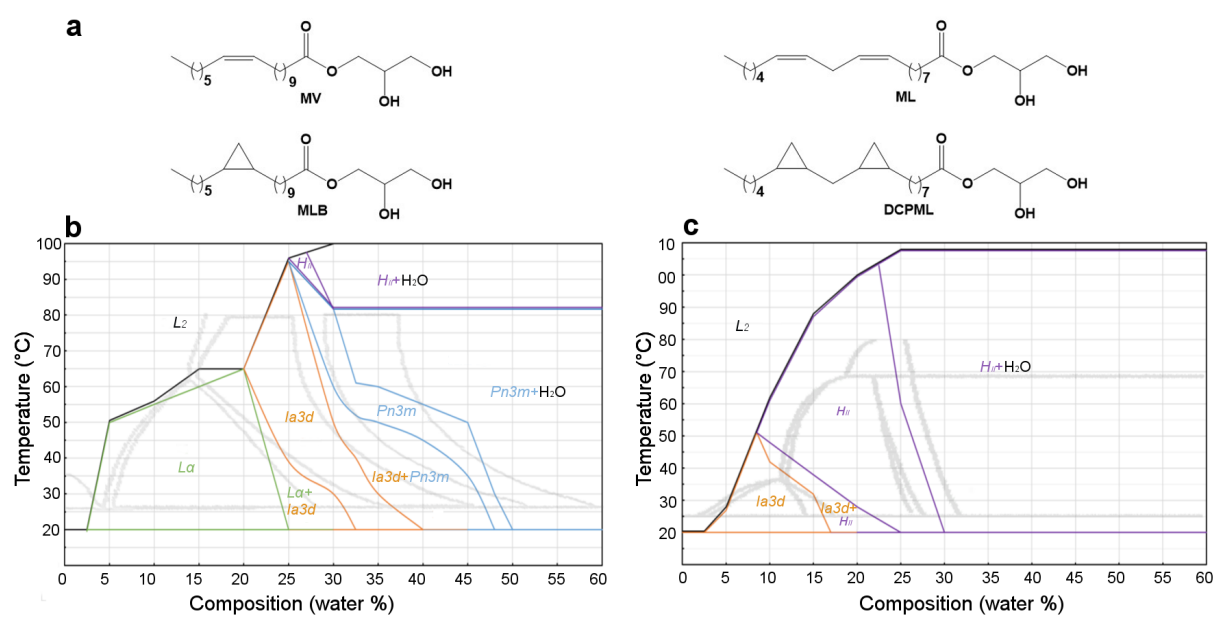

**Figure 1. Phase diagrams of cyclopropanated monoacylglycerols, MLB and DCPML a)** Chemical structure of the two olefinic monoacylglycerols MV and ML and of their cyclopropanated homologues MLB and DCPML. **b)** Sample composition/temperature phase diagram of the MLB:water system (solid lines) overlaid on MV: water phase diagram (background) adapted from Kulkarni *et al.*[18] **c)** Sample composition/temperature phase diagram of the DCPML: water (solid lines) overlaid on the phase diagrams of ML:water adapted from Kulkarni *et al.*[18]

Hydrated MLB presents a transition sequence similar to that of classic monoacylglycerols (**Fig. 1b**), where $L_α$, *Ia3d*, and *Pn3m* are observed upon increasing levels of hydration. In contrast to MDS,[28] the highly curved $H_{II}$ phase is present in MLB at high temperature. The $H_{II}$ phase and the *Pn3m* cubic phase were found to be stable in excess of water, and it was possible to determine a full hydration line for both phases through the analysis of the lattice parameters at each level of hydration.

In the case of the most curved lipid in the series, DCPML, a most remarkable and unusual phase behaviour was observed (**Fig. 1c**): the *Ia3d* cubic phase readily formed at 22°C and water content as low as 5% (w/w). The phase sequence upon hydration from *Ia3d* to $H_{II}$ was markedly different from the most common sequence of phases for monoacylglycerols ($L_α$ → *Ia3d* → *Pn3m*). Previous studies have shown that pure hydrated monoacylglycerols can form $H_{II}$ only at high temperatures. Stable $H_{II}$ phases at room or physiological temperature require addition of hydrophobic molecules,[32,33] or the presence of an ether rather than an ester bond linking the head group to the lipidic tail.[34] The characteristic slow molecular release from $H_{II}$ makes this geometry particularly appealing for several applications, such as drug delivery[35,36] and membrane proteins reconstitution,[37] but the complexity of ternary systems has limited the popularity of this mesophase.

Formation of the $H_{II}$ phase at room temperature suggests a shift of the DCPML phase diagram to lower temperatures and hydration, which was confirmed by a dedicated investigation. A DCPML sample with 12.5% w/w of water was first cooled in a stepwise manner to -20°C, then heated back to 22 °C. At the end of each step in both directions, the system was equilibrated and SAXS spectra were collected (**Fig. 2a**). The phase transition from $L_α$ to *Ia3d* occurs between -15 °C and -10 °C in both the heating and cooling directions, and reveals a novel stable lipidic cubic phase at sub-zero temperatures. As expected, the water channel radius of the *Ia3d* phase decreases upon heating from a maximum of 8.4 Å at -10°C to 7.8 Å at 22°C (see SI). The stability of the *Ia3d* cubic phase at sub-zero temperatures is remarkable, and is in contradistinction to all the other lipids that are known to form cubic phases, such as MO, which crystallize in a lamellar crystalline phase and ice below 0 °C.[17]

**Water behaviour**

The liquid crystal nature of DCPML at sub-zero temperatures suggests nontrivial features of the water confined in the nanochannels. The size of the water regions (slabs or channels) can be tuned by changing the water/lipid ratio (see SI). Melting transitions were studied by DSC measurements of mesophases at various hydration levels (**Fig. 2b**). Samples of DCPML were treated thermally in a heat/cool/heat cycle from -70°C to 60°C with a scanning rate of 5 °C/min. The traces presented in the figure were obtained during the second heating process. A sharp ice melting peak at 0°C, characteristic for pure water, appears in the samples with 25% and 20% w/w of water. As hydration is lowered, the peak first broadens (15% w/w of water) and then completely disappears in the samples with 5% and 10% w/w of water. Therefore, confinement within the $L_α$ and *Ia3d* phases in the low hydration regime is strong enough to completely prevent water crystallization at the considered cooling rate. This result is in line with previous work, where the impossibility of ice nucleation under extreme confinement has been ascribed to the need of formation of a minimal ice nucleus of radius 1 nm.[38] Moreover, confinement effects have been reported in the range between 1 to 100 nm,[9] in addition to extremely reduced mobility in pores with diameters between 4 and 50Å.[39] The small peaks observed at higher temperatures in **Fig.2b** correspond to the transitions between different geometries, and are in line with SAXS results (**Fig.1c**). The deviation of a few degrees in the transition temperatures can be explained by different heating rate and subsequent equilibration time, as well as an error (±1.5%) in the sample composition (water) considering different batches of lipids. The absence of ice for temperatures as low as -60 °C is a unique feature of DCPML: monolinolein at similar compositions always shows the transition peak for ice formation.

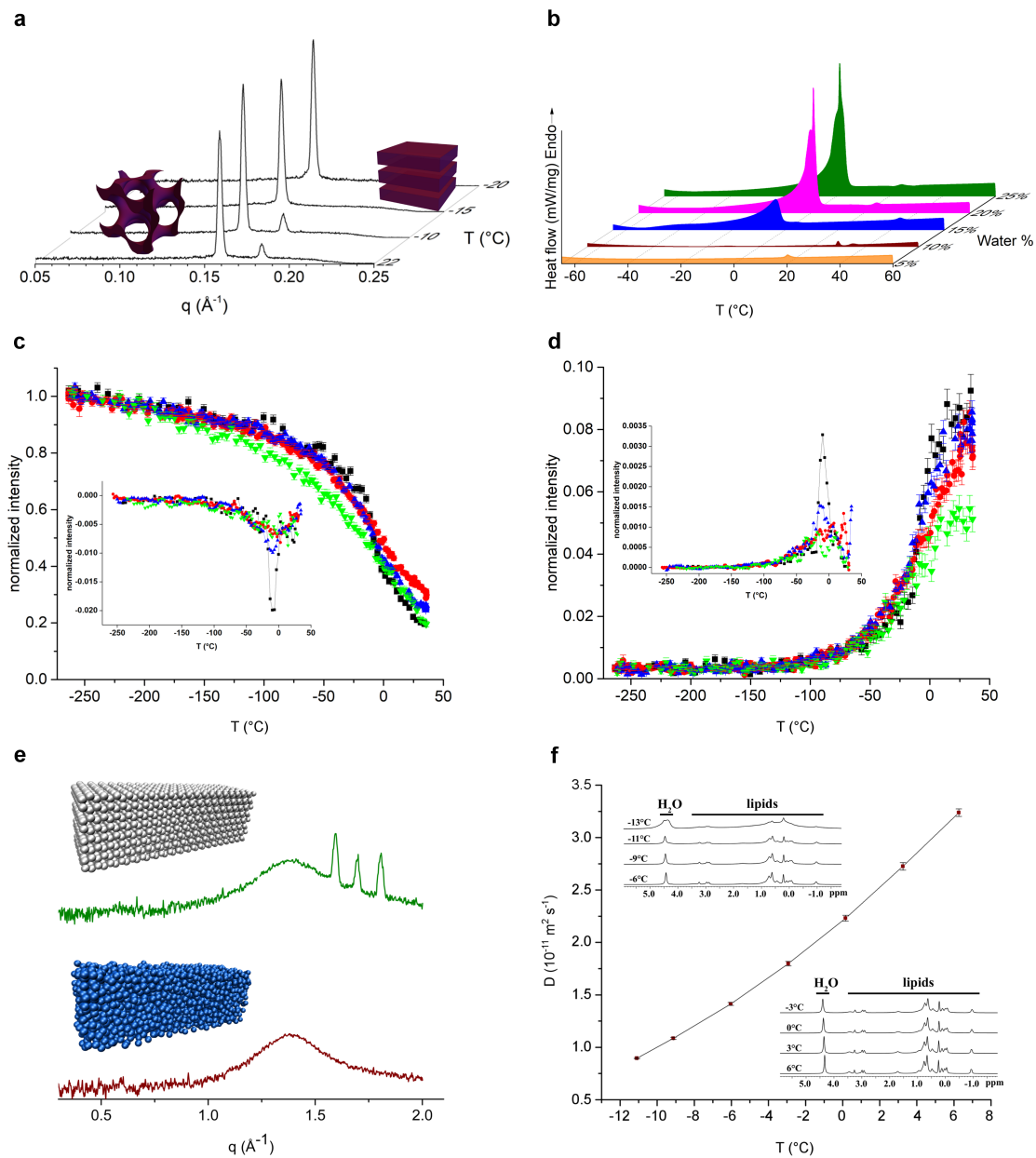

**Figure 2. Low temperature phase studies of water confined in DCPML mesophases a)** SAXS profiles of DCPML mesophases containing 12% (w/w) water at selected temperatures. Phase transition from $L_\alpha$ to $Ia3d$ occurs between -15 and -10 °C. **b)** DSC traces of DCPML with water content ranging from 5 to 25% w/w. The peaks for water crystallization occur in the samples with 25% (olive green), 20% (magenta), 15% (blue) w/w water, and are absent in the samples with 10% (dark red) and 5% (orange) w/w water. **c), d)** Temperature dependent neutron elastic **c)** and inelastic **d)** scattering of the DCPML samples with 15% (blue), 7.5% (red) w/w water, 15% w/w $D_2O$ (light green), and ML 7.5% w/w water (black). The inset shows the derivative of the scattering intensity. DCPML sample with 7.5% w/w water (red) and 15% w/w $D_2O$ (light green) do not show any transition peak for water freezing, while the DCPML samples with 15% w/w water (blue) and ML with 7.5% w/w water (black) show transition peaks at -9°C and -8°C, respectively. **e)** WAXS spectra at -30 °C of a DCPML sample with 10% (dark red) and 25% (olive green) w/w water. Crystalline scattering pattern is absent in the sample with 10% w/w water, while a profile typical of hexagonal ice is obtained from the sample with 25% w/w water. The broad peak present in both profiles is characteristic for the $L_\alpha$ phase. **f)** Temperature dependence of the water diffusion coefficient in the sample with 10% w/w water estimated by NMR. NMR spectra at different temperatures are reported in the inset.

FWS measurements (**Fig. 2c, 2d**) were employed to confirm the absence of a first order transition in DCPML at low water content. In bulk water, the mobility changes abruptly at 0 °C, resulting in a sudden jump in the neutron scattering intensity. The samples were cooled to -263 °C at a rate of 0.1 °C min$^{-1}$, equilibrated, and subsequently heated at the same rate. Measurements were performed during the heating cycle. The sample with 7.5% w/w of water was chosen to ensure a single geometry in the sub-zero temperature range. The FWS profiles show no jump around 0 °C, although a change of slope indicates increased water mobility starting from around -50°C. In contrast, when a mesophase with the same topology and water content is produced from the commercial ML instead of DCPML, water melting is clearly observed around -10°C, as evidenced by a sharp peak in the first derivatives (**Fig. 2c, 2d inset**). DCPML with 15% w/w of water also shows a jump corresponding to ice melting at around -10°C. Nevertheless, in this case the intensity of the transition is smaller, indicating that only part of the water is involved in ice formation, in agreement with DSC. Moreover, samples hydrated with $D_2O$ instead of $H_2O$ were employed to isolate the signal of the protonated lipid molecules, i.e. the lipidic contribution to the overall mobility of the system. The lack of a jump for a lipid-lipid transition confirms the absence of the $L_c$ crystalline phase in DCPML (**Fig. 2c-d**).

WAXS experiments at low temperature reveal a hexagonal structure of the ice in the samples with 20% and 25% w/w of water (**Fig. 2e**), as well as the absence of a crystalline arrangement in the WAXS region for the other samples, thus further confirming the liquid crystalline nature of the lamellar phase ($L_\alpha$) and, for low hydration, the absence of crystalline ice.

Water mobility and phase behaviour were also investigated by diffusion NMR (**Fig. 2f**). For the sample with 7.5% w/w, the detection limit was reached at 0 °C, indicating a diffusion coefficient $D$ lower than $10^{-11}$ m$^2$/s (the smallest value measurable with our NMR setup), whilst for the sample with 10% w/w diffusion was observed until -11 °C. The quasi-linear dependence of $D$ on temperature (**Fig. 2f**) confirms the liquid nature of water in the considered range of temperatures, and together with the FWS and DSC results indicates a transition from liquid to glassy water at lower temperatures.

**Lipid and water behaviour dependence**

By combining the data acquired by DSC, FWS, WAXS and NMR, the phase diagram of water confined in DCPML mesophases was established (**Fig. 3a**). The enthalpy peaks in the DSC and the intensity jumps in FWS were employed to draw the line at which water crystallizes into hexagonal ice as confirmed by WAXS. The transition between liquid and glassy water was determined by combination of NMR and WAXS, respectively addressing low mobility and amorphous structure of the water phase.

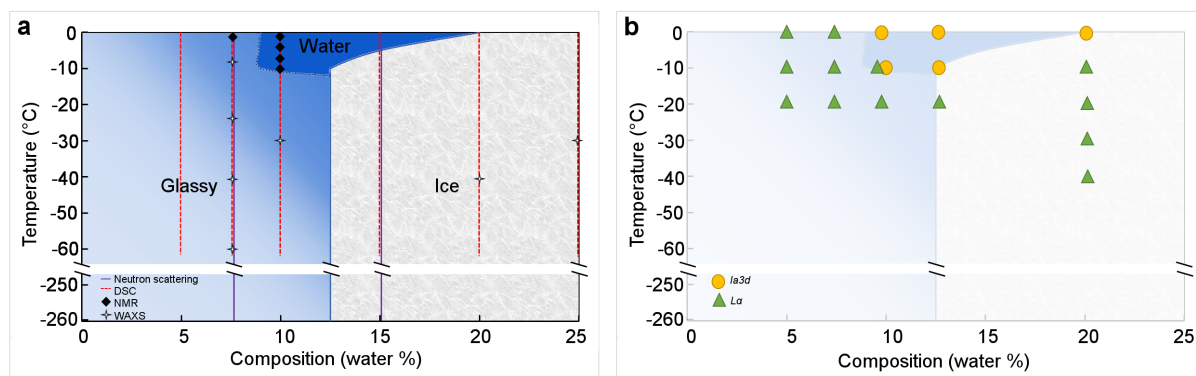

**Figure 3. Low temperature phase diagram of the water and DCPML lipidic phase a)** Sub-zero phase diagram of $H_2O$ confined in DCPML mesophases based on DSC, neutron scattering, WAXS and NMR data. **b)** Sub-zero phase behaviour of DCPML determined by SAXS measurements overlaid on the phase diagram of the confined water.

These features are intimately related to the peculiarities which distinguish DCPML from all the other known monoacylglycerols, namely a general shift of the phase transitions to lower temperatures and hydration, and the absence of $L_c$ even at extremely low temperatures. The steric hindrance of the cyclopropyl groups, and the presence of several diastereomers that differ in the relative configurations of the two *cis*-cyclopropyl groups, are understood to be the reasons preventing the packing of the lipidic chain of DCPML in a crystalline phase.

The lipidic geometry at low-temperature was revealed by systematic SAXS measurements, and results are shown in **Fig. 3b** overlaid on the water phase diagram. Upon hydration, a reentrant transition $L_\alpha \rightarrow Ia3d \rightarrow L_\alpha$ can be observed in the temperature range between -10° C and 0°C. Remarkably, the presence of liquid water at sub-zero temperatures is associated with the stability of the $Ia3d$ cubic phase. When cooling down at low hydration, the combination of lipid disorder and geometrical confinement of the $L_\alpha$ phase (e.g. DCPML with 7.5% w/w of water at 0°C has water slabs thickness of 3 Å) prevents ice formation at any temperature.

By increasing hydration, only part of the water molecules are bound to the lipid-water interface, resulting in freezing of the "bulk-like" core into hexagonal ice upon cooling. Upon cooling the sample with 20% w/w of water until -30°C, SAXS and WAXS showed that the $Ia3d$ phase is stable for hours, with no ice detected. The transition to $L_\alpha$ with a water layer thickness of 8.2 Å occurs after incubation for 1 h at -40°C, together with the appearance of ice peaks in the WAXS region. Upon heating from -40°C, the $L_\alpha$ phase is stable until 0°C. Between -40°C and -20°C, the lattice parameter $a$ shows the typical decrease observed for mesophases (from 39.2 Å to 38.4 Å). In contrast, $a$ increases between -20°C (38.4 Å) and -10°C (39.2 Å), which is usually associated with increased hydration of the lipid bilayer. Remarkably, this temperature range corresponds to the onset of ice melting observed by DSC (**Fig. 2b**), thus correlating the swelling to the progressive rehydration of the $L_\alpha$ phase. Upon transition from $L_\alpha$ to $Ia3d$ ($a$= 110.2 Å) at 0°C there is no evidence of hexagonal ice. Although metastability of the lipidic phase has been observed in monoacylglycerols,[17] the extraordinary hysteresis between cooling and heating in the samples in which ice is formed, together with the reentrant lipidic phase behaviour, suggests a competition between the formation of ice and the stability of the cubic phase.

## Molecular Dynamics simulations

Our experimental findings were further supported by Molecular Dynamics simulations. To that end, a systematic study of ice melting and water mobility within $L_\alpha$ and $Ia3d$ geometries was performed.

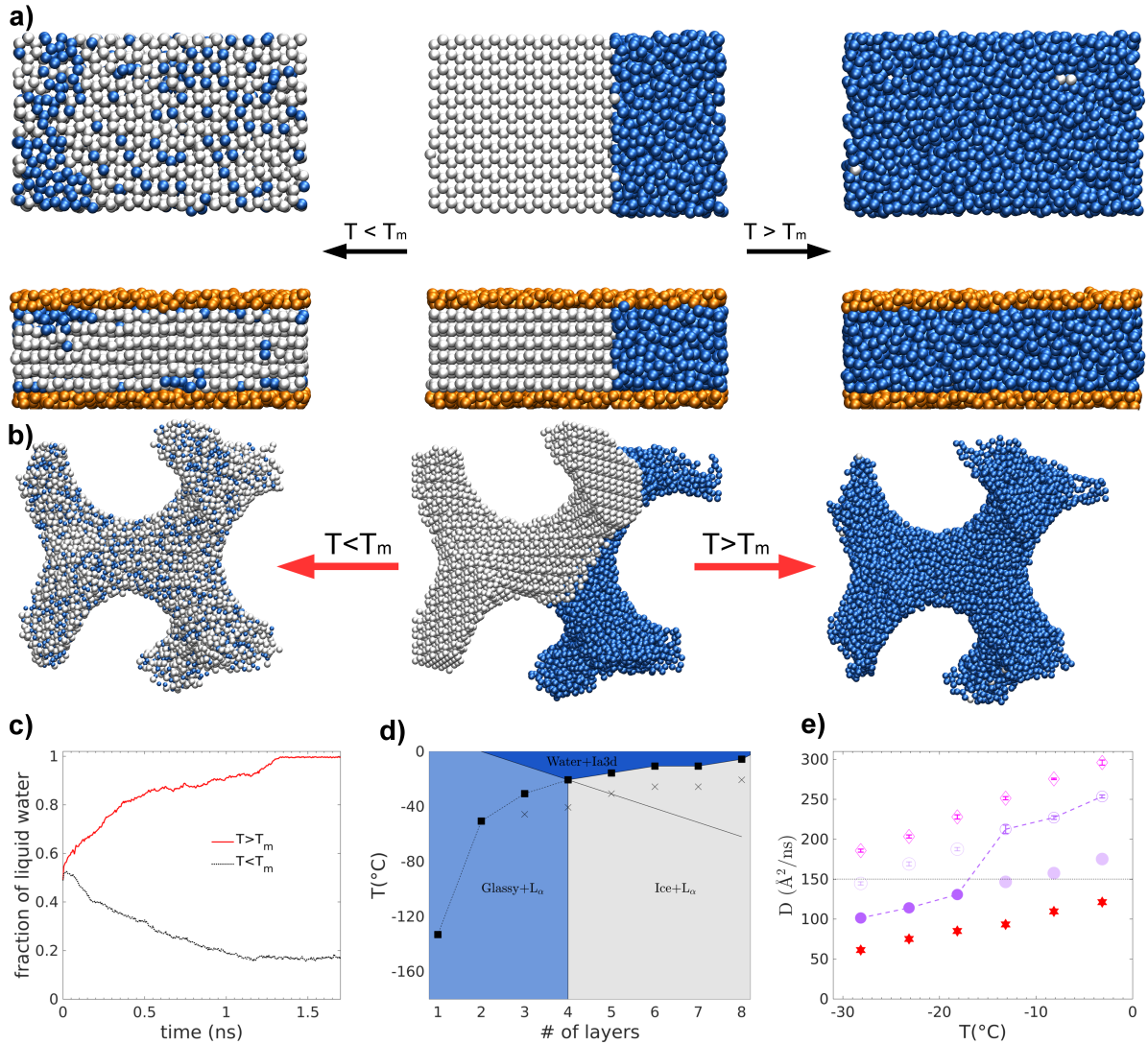

**Figure 4. MD simulations of water and ice under confinement a)** Representative snapshots of MD simulations on the melting point of lamellar mesophases for *n*=6 ice layers. Centre: starting configuration, partially filled with ice (white beads) and liquid water (blue beads). Left: final configuration below the melting point. Right: final configuration above the melting point. Top and bottom row give perspective orthogonal and parallel to the hexagonal plane, respectively. In the bottom row, lipid-like molecules are depicted as orange beads. **b)** Representative snapshots of water confined within an *Ia3d* cubic phase with radius 17.3 Å, for starting (centre) and final configurations below (left) and above (right) the melting point. To enhance visibility, lipid-like molecules are not drawn. **c)** Time evolution of the fraction of liquid water for the same system as in a), above (red continuous line) and below (black dotted line) the melting point. **d)** Simulated phase diagram obtained combining the results for $T_m$ within $L_\alpha$ (squares and dashed line) and $Ia3d$ (crosses), and dependence of water mobility upon change in mesophase geometry (continuous line). **e)** Diffusion coefficient vs temperature for $L_\alpha$ with 2-layer thickness (full red stars); $L_\alpha$ with 4-layer thickness (full purple circles); and the corresponding *Ia3d* with the same amount of water (radius *r*=13.9 Å, empty purple circles); *Ia3d* with radius 19.1 Å (empty magenta diamonds, same amount of water as $L_\alpha$ with 7-layers thickness). The dashed purple line shows the path followed upon cooling by the system with the composition denoted by circles (4-layer $L_\alpha$ and *Ia3d* with *r*=13.9Å), indicating the sudden drop of *D* due to the lipid-lipid transition. The horizontal grey line denotes the chosen threshold to consider a system as glassy.

We considered the coarse-grained mW force field,[40] which has been successfully employed to address the features of supercooled water[41] and ice melting in cylindrical nanotubes[38,42] (see Materials and Methods). The lipid-water interface was represented as liquid water molecules forced to keep a disordered configuration at any temperature via harmonic springs.[38] This choice reflects the unique lipidic liquid crystalline nature of DCPML at low temperatures, and ensures the hydrophilicity of the interface.[42] Subsequently, the water domains were filled with liquid water and hexagonal ice in contact with each other (**Fig.4a** and **Fig.4b**). This setup enables computation of the ice melting point $T_m$ by systematic simulations at different temperatures $T$:[43] If $T<T_m$, the ice nucleus propagates to the entire system, while for $T>T_m$ it melts (**Fig. 4**). $T_m$ for hexagonal ice at various degrees of confinement was computed, namely considering $L_\alpha$ water slabs containing 1 to 8 ice layers, and the corresponding *Ia3d* geometries with the same water content (see SI). Thus, the melting point under confinement is generally lower than in bulk (**Fig.4d**), in direct agreement with the DSC experiments. For a given geometry, $T_m$ increases with the size of the confining region, in accord with previous reports on different systems[44]. When comparing $L_\alpha$ and *Ia3d* geometries with similar water content, $T_m$ is systematically lower in the latter case, thus pointing to the existence of a temperature range where only $L_\alpha$ is compatible with the presence of ice.

A second set of simulations addressed the experimental observation of glassy water upon transition from cubic to lamellar geometries, by studying the mobility of liquid water inside *Ia3d* and $L_\alpha$ mesophases. To this aim, a system containing only liquid water was quenched at temperatures between -30 and 0 °C. After a short relaxation, the average self-diffusion coefficient $D$ of the molecules in the aqueous phase was computed, obtaining the plots reported in **Fig.4e**. The horizontal grey line indicates the liquid/glassy transition, whose value was arbitrarily fixed to 150 Å$^2$/ns for explanation purposes. This value is unusually large for real water, and is a consequence of the known overestimation of water diffusion by the Mw model.[40] Three important features emerge from the results. First, for a given system, $D$ increases as a function of temperature and shows a linear behaviour in the considered range. Second, for both $L_\alpha$ and *Ia3d*, at any temperature the diffusion coefficient is smaller for more confined water phases, in agreement with previous experimental and theoretical observations.[15,45] Third, mobility within *Ia3d* is systematically larger than in $L_\alpha$ with the same hydration. These results indicate that the lipid-lipid transition has a major impact on diffusion, thus explaining the simultaneous observation of liquid-to-glass transition in correspondence of the *Ia3d*→$L_\alpha$ order-order transition detected experimentally at low hydration (**Fig.3**). Following this argument, in **Fig.4d** an indicative lipid-lipid transition curve (continuous line), depicted with a negative slope as explained by theory[46] and observed by experiments,[18] separates the liquid and glassy phases of water, due to the mobility drop induced by the lipid transition (**Fig.4e**).

The quantitative details of the results depend on the particular choices made, *i.e.* simplified force field, water-like interaction between lipid and water molecules, lipid-lipid transition line, and glass threshold. Nevertheless, their qualitative features are very robust and corroborate the experimental observations, indicating the following physical picture. At low hydration, the system follows the "standard" behaviour, *i.e.* a transition line from cubic to lamellar with negative slope (**Fig.3** and **Fig.4d**).[18,46] Upon cooling, the *Ia3d* phase switches to $L_\alpha$, experiencing a sudden drop in water mobility (**Fig.4e**). Lower mobility implies that the system requires more time to equilibrate. Therefore, in this regime the cooling process crosses the melting line (**Fig.4d**) after diffusion has already been hampered, *i.e.* prior to water crystallization, thus resulting in the observation of glassy water (**Fig.3** and **Fig.4d**). At a critical hydration, the lipid-lipid transition line crosses the lamellar ice-melting line. Upon further cooling beyond this point, the *Ia3d* mesophase with highly-mobile water enters a region of the phase diagram where freezing is possible, but only within a lamellar arrangement. This induces a forced structural transition, thus leading to a lipid-lipid transition which follows the $T_m$ lamellar ice-melting line with positive slope. These two different slopes are at the origin of the $L_\alpha \rightarrow Ia3d \rightarrow L_\alpha$ reentrant behaviour. This mechanism can be ascribed to the large difference in the enthalpy change of lipid-lipid and ice-water phase transitions, as evidenced by the DSC curves. For example, in the 20% profile (**Fig.2b**), computation of the peak area at ~ 0°C (ice-water transition) and at ~ 30°C (lipid-lipid transition) yielded an 85-fold difference. Due to the large enthalpy difference, the system finds it thermodynamically convenient to switch geometry and let water freeze, although the barriers involved in such a major rearrangement can easily lead to the observed hysteresis.

In summary, we designed, synthesized and investigated a new library of monoacylglycerol-type lipids in which the naturally occurring *cis* double bonds were substituted by rigid cyclopropyl moieties, thus preventing lipid crystallization. In this way, the rich polymorphism of lipidic mesophases is extended for the first time to sub-zero temperatures, opening unexplored possibilities in the storage of amorphous water within soft biological interfaces. The most remarkable features enabled by such lipid molecular design are the presence of liquid water down to -10°C in the *Ia3d* double gyroid cubic phase, and glassy water down to -263 °C in the $L_\alpha$ lamellar phase. These results expand our understanding of how two main components of life, i.e. water and lipids, interact under extreme conditions of temperature and geometrical confinement. Beside their fundamental significance, our findings may also open new perspectives and opportunities in the use of such tailor-made lipidic mesophases in low-temperature biomimicry, as well as biochemical and biophysical investigations, with special emphasis on complex biomacromolecules that are unstable at room temperature.

**Materials and Methods**

Cyclopropanated lipids were synthesized according to the procedure previously reported by Salvati Manni et al.[28] For both lipids the synthesis was accomplished in two steps starting from the corresponding commercially available acid 1. Cyclo-propanation of the olefins was carried out with diethyl zinc and diiodomethane in the presence of 2,4,6-trichlorophenol at -40°C to afford *cis*-cyclopropaneted derivatives of the acids 2 in 69-71% yield; which were then reacted with an excess of glycerol in the presence of 1-Ethyl-3-(3-dimethylaminopropyl)carbodiimide (EDC) and 4-(dimethyl-amino)pyridine (DMAP) in a mixture of dichloromethane and dimethylformamide (DMF) to afford the desired ester 3 in 58-71% yield. The resulting synthesized monoacylglycerols are a mixture of all the possible enantiomers and diastereomers. Moreover the natural transesterification of the head group of monoacyglycerols creates an equilibrium between the 1- and 2- isomers. Purity of the lipids was confirmed by $^1$H-NMR, $^{13}$C-NMR and mass spectroscopy.

**Mesophase sample preparation**

Liquid-crystalline mesophases were prepared by mixing weighed quantities of desired lipid and water (to achieve the desired composition of each sample of the phase diagram) inside sealed Pyrex glass tubes, followed by melting to a fluid isotropic state by gentle heating, mixing by vortex and centrifugation until a homogenous mixture was obtained. The prepared mesophases were then allowed to cool to room temperature over a period of 24 h in order for them to reach a state of thermodynamic equilibrium before SAXS measurements were performed.

An alternative method using two 100 μL Hamilton syringes loaded with the weighed quantities of lipid and water, respectively was also employed. After homogenization with the Hamilton coupled syringe setup, the obtained mesophases were transferred to 2mm quartz glass capillaries (Hilgenberg) between two layers of Teflon and sealed with a solvent free 2-component epoxy resin adhesive (UHU plus). The sample was then melted to form an isotropic fluid, and subsequently equilibrated at room temperature for at least 24 h before measurements.

**Small-angle X-ray scattering (SAXS) measurements**

SAXS measurements were employed to determine phase identity, identify the symmetry and unit cell parameters of the mesophases at the different conditions. Experiments were performed on a Rigaku SAXS instrument with MicroMax-002+ microfocused beam, operating at voltage and filament current of 45 kV and 0.88 mA, respectively. The Ni-filtered Cu Kα radiation ($\lambda_{Cu\,K\alpha}$=1.5418 Å) was collimated by three pinhole (0.4, 0.3 and 0.8 mm) collimators and the data were collected with a two-dimensional argon-filled Triton detector. An effective scattering-vector range of 0.03 Å$^{-1}$<q<0.45 Å$^{-1}$ was probed, where $q$ is the scattering wave vector defined as $q=4\pi \sin(\theta)/\lambda_{Cu\,K\alpha}$, with a scattering angle of $2\theta$, calibrated using silver behenate. 2 mm quartz glass capillaries containing the sample were placed into a stainless steel holder.

Additional measurements were performed on a Bruker AXS Micro, with a microfocused X-ray source, operating at voltage and filament current of 50 kV and 1,000 μA, respectively. The Cu Kα radiation ($\lambda_{Cu\,K\alpha}$ = 1.5418 Å) was collimated by a 2D Kratky collimator, and the data were collected by a 2D Pilatus 100K detector.

An effective scattering-vector range of 0.04 Å$^{-1}$<q<0.5 Å$^{-1}$ was probed. Samples were placed inside a stainless steel cell between two thin replaceable mica sheets and sealed by an O-ring, with a sample volume of 10 μL and a thickness of ~1 mm. Measurements were performed at different temperatures, and samples were equilibrated for 30 minutes prior to measurements, while scattered intensity was collected over 30 minutes. For the low-temperature experiments, special care was taken to exclude supercooling: samples were cooled down at the rate of 1°C/min and equilibrated for 3 h at the lowest temperature. SAXS spectra were collected increasing the temperature after sample equilibration for 30 minutes at each temperature.

Mesophases were identified by their specific Bragg peak positions. For the double diamond cubic phase (*Pn3m*) the relative positions in $q$ of the Bragg reflections are at √2:√3:√4:√6:√8:√9…, whereas for the gyroid bicontinuous cubic phase (*Ia3d*) the Bragg peaks are at q = √6:√8:√14:√16:√20…. The H$_{II}$ phase is identified by reflections at 1:√3:√4…, the bilayer lamellar (*Lα*) phase exhibits Bragg peaks in the ratio of 1:2:3:4… and the L$_2$ phase is identified by a single characteristic broad peak. The mean lattice parameter, a, was deduced from the corresponding set of observed interplanar distances, d (d = 2π/q), using the appropriate scattering law for the phase structure. For cubic phases:

$$a = d\sqrt{h^2 + k^2 + l^2} \qquad (1)$$

while for the H$_2$ phase:

$$\frac{4\pi}{q\sqrt{3}}\sqrt{h^2 + hk + k^2} \qquad (2)$$

For the L$_2$ phase, which shows only one broad peak, d is termed the characteristic distance.

**Wide-angle X-ray scattering (WAXS) measurements**

WAXS measurements were employed to determine the water crystallinity at different hydration level and temperatures. Samples were prepared and handled as described for SAXS measurements.

Experiments were performed on a Rigaku SAXS instrument with MicroMax-002+ microfocused beam, operating at voltage and filament current of 45 kV and 0.88 mA, respectively. The Ni-filtered Cu Kα radiation ($\lambda_{Cu\,K\alpha}$=1.5418 Å) was collimated by three pinhole (0.4, 0.3 and 0.8 mm) collimators and the scattered X-ray intensity was detected by a Fuji Film BAS-MS 2025 imaging plate system. An effective scattering-vector range of 0.2 Å$^{-1}$<$q$<2 Å$^{-1}$ was probed.

**Differential scanning calorimetry (DSC)**

DSC experiments were performed on a Thermal Advantage DSC from TA Instruments. The samples were loaded in aluminium capsules contained in a steel sample pan of known mass and thermal conductivity. The experiments were carried out with an empty sample pan as the reference. Samples were scanned at a scanning rate of 5 °C/min in a heat/cool/heat cycle. They were first heated up from room temperature to 70 °C, then cooled down to -70, allowed to equilibrate for at least 48 h and subsequently scanned over a temperature range from −70 to 60 °C.

**Self-diffusion of H$_2$O by $^1$H-NMR**

Diffusion NMR was measured on a Bruker AV-III 500 MHz spectrometer equipped with a room temperature, z-gradient BBI probe-head using the stimulated echo pulse sequence with bipolar gradients and longitudinal eddy-current delay as found in the Bruker standard pulse-sequence library.[47] The diffusion time was set to (200 ms) and the eddy-current delay to 5 ms. The apparent diffusion coefficient at each temperature was obtained by non-linear least squares fits of the water signal intensities found for 16 different magnetic field gradient strength varying in a linear fashion between 2 and 98% of the maximum gradient strength (53.5 G/cm) to the relevant expression:

$$I(q) = I(0) \exp[-Dq^2(\Delta - \frac{\delta}{3})]$$

with $q = \gamma g \delta$. $\gamma$ is the gyromagnetic ratio, $g$ the gradient strength, $\delta$ the total gradient pulse width of a bipolar gradient pulse pair, $\Delta$ the diffusion time and $D$ the diffusion coefficient.

**Neutron spectroscopy**

Neutron spectroscopy was used to measure the portion of static Hydrogens in the sample (so called fixed window scan, FWS), and so as a complementary tool to detect phase transitions. The samples were cooled down from room temperature to -263 °C, while data were collected during heating back to room temperature at the rate of 0.1 °C/min. Neutrons scatter elastically on immobile atomic nuclei, while all type of motions, like vibrations or diffusion cause inelastic scattering of the neutrons. Therefore, at finite temperature the intensity of elastically scattered neutrons is reduced by the Debye-Waller factor, which accounts for thermal vibrations. Melting e.g. of bulk (like) water is a first order phase transition and would cause a sharp decrease of the intensity. The signal intensity is directly proportional to the number of H (other elements can be neglected for the present case). The sample is a mixture of lipids and water, which has nearly the same hydrogen density. The measurements have been performed at the MARS spectrometer (SINQ, Paul Scherrer Institut, Switzerland) using an energy resolution of about 13microeV. This corresponds to an observation time of about 100ps; motions which are slower are not detected. Simultaneously to the elastically scattered neutrons, also those with a small energy transfer (20-60 microeV) are recorded, which happens in case of diffusion. 150mg sample was distributed between two thin Aluminum sheets (15mm x 50mm) and enclosed in an Aluminum sample holder. The sample was perpendicular to the neutron beam. Data treatment (energy integration, background subtraction and normalisation) was performed using an in-house script available at the instrument.

**Molecular Dynamics simulations**

Simulations of the system were performed by means of the mW model of water.[40] In this force field, water is modelled as a four-valence element with intermediate properties between carbon and silicon, by means of a modified Stillinger-Web potential. Each molecule is represented as a single bead subjected to two- and three-body potentials. We refer the reader to Molinero et al.[40] for the exact formulation of the force field and the values of the parameters. As for the lipids, following Moore et al.[38] we considered a configuration of liquid water within the region corresponding to the lipid bilayer (a simple slab or a region surrounding the minimal surface of a gyroid for lamellar and *Ia3d*, respectively) and forced the system to keep such configuration via harmonic constraints with elastic constant equal to 30 Kcal/mol. The region corresponding to the lipid bilayer in the *Ia3d* case was identified by means of the algorithm proposed by Assenza et al.[45] All the molecules in the system interacted via the mW potential. The simulations were run in the NVT ensemble via a Nosé-Hoover thermostat with relaxation time 1 ps and integration step 10 fs.

To find the melting point, we adopted the approach from García Fernández et al.,[43] by putting in contact a nucleus of ice and liquid water. Based on the WAXS measurements, we built the ice nucleus according to the hexagonal structure. In the case of the lamellar geometry, we chose to put the hexagonal planes parallel to the lipid-water interface, as it is the configuration which is expected to be most stable. In the case of *Ia3d*, little influence is expected from the precise orientation due to the large orientational heterogeneity of the channels. The amount of ice and liquid water was monitored in time by means of the CHILL algorithm,[38] with the exception of the monolayer case, where we considered the trigonal order parameter $q_3$ from Henchman et al.,[48] and assigned to the liquid phase the molecules satisfying $q_3<0.995$. For each geometry and temperature, we run simulations up to 10 ns. If in the majority of simulations the final number of molecules assigned to the liquid phase was larger than in the starting configurations, the system was considered to be above the melting point; otherwise, it was below. As we proceeded with temperature steps equal to 5°C, the final melting point has an indeterminacy of 2.5°C.[38]

Diffusion simulations were prepared in a similar manner, although in this case the slab was simply filled with liquid water. For each temperature, the system was quenched and shortly equilibrated (0.1 ns). Then, the average mean-square displacement *MSD(t)* was computed, considering either all three dimensions (for *Ia3d*) or only the two dimensions along the slab (for lamellar). The diffusion coefficient *D* was computed by fitting *MSD(t)=4Dt* for lamellar geometries or *MSD(t)=6Dt* for *Ia3d*. In the latter case, care was taken in order to consider mean-square displacements short enough to ensure that the labyrinthic features of the network of water channels do not affect the computation.[45]

**Acknowledgements**

We acknowledge Prof. Jay Siegel for insights and useful discussions, and Dr. Wye Khay Fong for assistance with SAXS experiments. Support by SNF Sinergia grant CRSII2_154451 to EML and RM is gratefully acknowledged.

**Additional Information**

†LSM and SA contributed equally to this work. Correspondence should be addressed to EML (ehud.landau@chem.uzh.ch) and RM (raffaele.mezzenga@hest.ethz.ch).